\documentclass[traditabstract]{aa}
\bibpunct{(}{)}{;}{a}{}{,} 
\usepackage{graphicx}
\usepackage[varg]{txfonts}
\usepackage{color}

\usepackage[switch,pagewise]{lineno}
\begin{document}

\title{Spatially separated continuum sources revealed by microlensing in the gravitationally lensed broad absorption line quasar SDSS~J081830.46+060138.0\thanks{Based on observations made with ESO Very Large Telescope at the Paranal Observatory under program ID~100.B-0590. The reduced spectra are available at the CDS via anonymous ftp to http://cdsarc.u-strasbg.fr.}}
\author{D. Hutsem\'ekers\inst{1,}\thanks{Senior Research Associate F.R.S.-FNRS},
        D. Sluse\inst{1},
        P. Kumar\inst{1,2}
        }
\institute{
    Institut d'Astrophysique et de G\'eophysique, Universit\'e de Li\`ege, All\'ee du 6 Ao\^ut 19c, B5c, 4000 Li\`ege, Belgium
    \and 
    Aryabhatta Research Institute of Observational Sciences (ARIES), Nainital, Uttarakhand, India
    }
\date{Received ; accepted: }
\titlerunning{Microlensing in quasar SDSS~J081830.46+060138.0} 
\authorrunning{D. Hutsem\'ekers et al.}
\abstract{Gravitational microlensing is a powerful tool for probing the inner structure of distant quasars. In this context, we have obtained spectropolarimetric observations of the two images of the broad absorption line (BAL) quasar SDSS~J081830.46+060138.0 (J0818+0601) at redshift $z \simeq$ 2.35. We first show that J0818+0601 is actually gravitationally lensed, and not a binary quasar. A strong absorption system detected at $z$ = 1.0065$\pm$0.0002 is possibly due to the lensing galaxy. Microlensing is observed in one image and it magnifies the emission lines, the continuum, and the BALs differently. By disentangling the part of the spectrum that is microlensed from the part that is not microlensed, we unveil two sources of continuum that must be spatially separated: a compact one, which is microlensed, and an extended one, which is not microlensed and contributes to two thirds of the total continuum emission.  J0818+0601 is the second BAL quasar in which an extended source of rest-frame ultraviolet continuum is found. We also find that the images are differently polarized, suggesting that the two continua might be differently polarized. Our analysis provides constraints on the BAL flow. In particular, we find that the outflow is seen with a nonzero onset velocity, and stratified according to ionization.
}
\maketitle
%
%
%
\section{Introduction}
\label{sec:intro}

Gravitational microlensing by compact objects in lensing galaxies is a powerful tool for probing the structure of distant quasars on sub-parsec scales \citep[e.g.,][for a review]{2010Schmidt}. Indeed, microlensing selectively magnifies the different regions that constitute the quasar core depending on their size. Usually taking place in one image of a lensed quasar, magnification of the compact accretion disk results in a flux variation of the ultraviolet (UV)-visible continuum, while weaker, partial microlensing of the more extended broad emission line (BEL) region (BELR) may induce line profile deformations. These variations occur on timescales of weeks to years due to the relative motions of the source, lens, and observer and they can provide constraints on the size, location, geometry, and kinematics of the quasar inner regions \citep[e.g.,][]{2008Eigenbrod,2011Blackburne,2007Sluse,2011Sluse,2011Odowd,2013Guerras,2016Braibant,2019Hutsemekers}.

Broad absorption line (BAL) quasars are characterized by deep blueshifted absorption in the resonance lines of ionized species, revealing fast and massive outflows \citep[e.g.,][]{2019Hamann}.  BALs are observed in approximately 20\% of optically selected quasars \citep[e.g.,][]{2008Knigge}. A detailed study of microlensing in the lensed BAL quasar H1413+117 provided interesting constraints on the outflow. In particular, the \ion{C}{iv}~$\lambda 1550$ BAL profile was found to consist of a completely black absorption which starts at an onset velocity of $\sim$2000 km~s$^{-1}$ and is partially filled in by the broad emission line, thus suggesting disk-wind or disk-wind and polar outflows \citep{2010Hutsemekers,2015Odowd}. An extended source of continuum was uncovered in addition to the compact, microlensed one \citep{2015Sluse}. Polarization microlensing suggested that the continuum scatters off two spatially separated regions that produce roughly perpendicular polarizations \citep{2015Hutsemekers}. These results demonstrate the potential of microlensing to probe outflows in BAL quasars. However, so far, a detailed microlensing analysis has only been carried out for H1413+117.

SDSS~J081830.46+060138.0 (hereafter J0818+0601) is a BAL quasar with a moderate \ion{C}{iv} balnicity index of 330 km~s$^{-1}$ \citep{2017Paris}. The redshift measured from the \ion{Mg}{ii} BEL is $z$ = 2.3669 \citep{2017Paris}. When looking at the spectrum, it appeared that the \ion{Mg}{ii} BEL was contaminated by telluric lines. Therefore, we adopt hereafter the systemic redshift $z$ = 2.349$\pm$0.003 that we measured from CIII] $\lambda 1909$, which is in better agreement with the position of the other emission lines. In a systematic search for new gravitationally lensed quasars, \citet{2016More} found that J0818+0601 has two components separated by about 1\farcs1, showing similar spectra. They classified it as a binary quasar mainly due to the nondetection of the lensing galaxy. Based on new observations presented in Sect.~\ref{sec:obs}, we argue that J0818+0601 is actually a two-image gravitationally lensed quasar (Sect.~\ref{sec:glens}). Since one image is affected by microlensing, we disentangle the microlensed and nonmicrolensed parts of the spectrum and compare the polarization of the two images in Sect.~\ref{sec:mulens}. Consequences for the quasar structure are discussed in Sect.~\ref{sec:discu}, while conclusions are found in the last section.

\begin{figure*}
\centering
\resizebox{\hsize}{!}{\includegraphics*{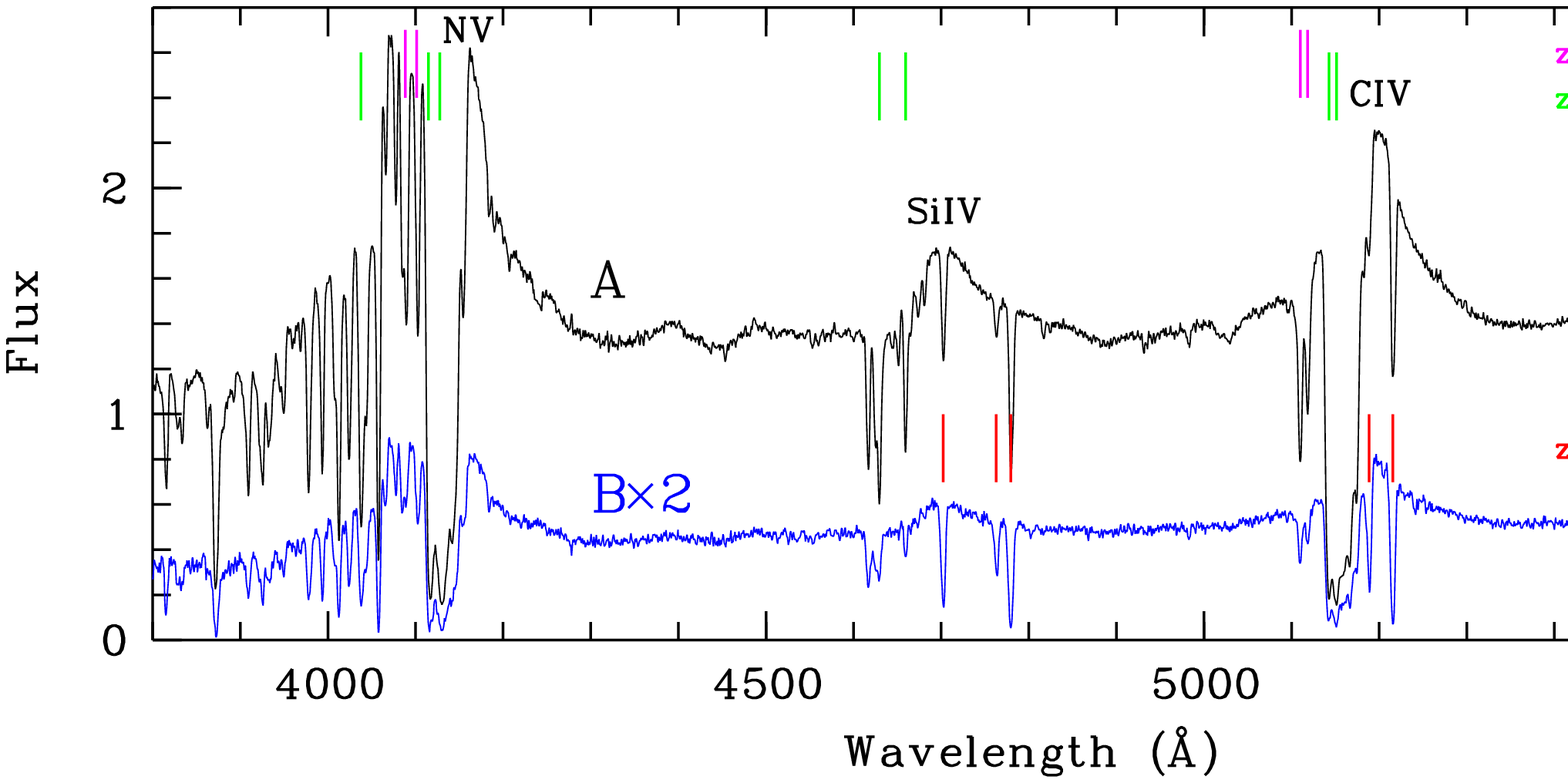}}\\
\resizebox{\hsize}{!}{\includegraphics*{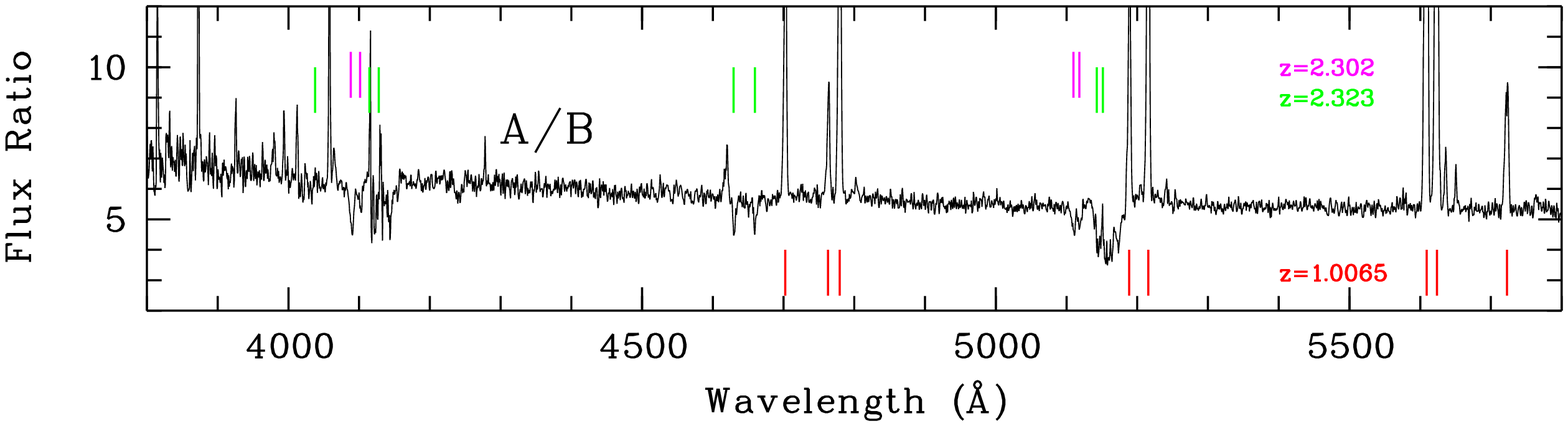}}
\caption{{\it Upper panel:} Spectra of images A (top) and B (bottom) of J0818+0601. To enhance visibility, the spectra have been divided by an approximate continuum identical for images A and B, and the spectrum of image~B has been multiplied by a factor 2. The spectra are given in arbitrary units. Green and magenta tick marks on the top indicate intrinsic absorption features at redshift $z$ = 2.323 (\ion{C}{iv}~$\lambda\lambda$1548.203,1550.777; \ion{Si}{iv}~$\lambda\lambda$1393.755,1402.770; \ion{N}{v}~$\lambda\lambda$1238.821,1242.804; Ly$\alpha$~$\lambda$ 1215.670) and $z$ = 2.302 (\ion{C}{iv}; \ion{N}{v}), respectively. Red tick marks in the middle indicate intervening absorption lines at redshift $z$ = 1.0065 identified in both images A and B (Table\ref{tab:lines}). {\it Lower panel :} A/B flux density ratio. Tick marks indicate the same features as in the upper panel.}
\label{fig:spectra}
\end{figure*}

\section{Observations and data processing}
\label{sec:obs}

Spectropolarimetric observations of the two images of J0818+0601 were obtained on December 24 and 25, 2017, using the European Southern Observatory Very Large Telescope (VLT) equipped with the FORS2 multimode instrument (run 100.B-0590), under excellent seeing conditions ($\sim$0\farcs6). The two images, separated by about 1\farcs1, were positioned in the same 0\farcs7-wide multiobject spectroscopy (MOS) slit located at the center of the field after moving the field rotator to the position angle -54.91\degr. The pixel size along the slit corresponds to 0\farcs125 on the sky.

Linear spectropolarimetry was performed by inserting a Wollaston prism in the beam that splits the incoming light rays into two orthogonally polarized beams. Spectra were secured with grism 600B so as to cover the spectral range 3500-6200~\AA\ with an average resolving power R$\simeq$1110. Because the two orthogonally polarized images of the object were recorded simultaneously, the polarization measurements are essentially independent on variable atmospheric transparency and seeing. In order to derive the normalized Stokes parameters $q$ and $u$, two observing blocks of four frames were obtained with the halfwave plate rotated at four different position angles, 0\degr, 22.5\degr, 45\degr, and 67.5\degr. This combination allowed us to remove most of the instrumental polarization. For each image of J0818+0601, 16 spectra (two orthogonal polarizations, four halfwave plate angles, and two epochs) were obtained and extracted by fitting two Moffat profiles along the slit (see Appendix \ref{sec:apa} for details). The normalized Stokes parameters $q(\lambda)$ and $u(\lambda)$, the linear polarization degree $p(\lambda)$, the polarization position angle $\theta(\lambda)$, and the total flux density $F(\lambda)$ were then computed from the individual spectra according to standard formulae \citep[e.g.,][]{2018Hutsemekers}.  In order to express $\theta(\lambda)$ with respect to the N-S direction as usual, the spectra were corrected for the rotator angle and for the retarder plate zero angle given in the FORS2 user manual. A comparison of the spectra and polarization measurements obtained on December 24 and 25, 2017 revealed no significant differences. Therefore, we only consider the data averaged over the two epochs throughout this paper.

\section{J0818+0601: a lensed BAL quasar}
\label{sec:glens}

Fig.\ref{fig:spectra} shows the spectra of components A and B of J0818+0601 as well as the A/B flux density ratio. Two systems of intrinsic absorbers are detected. Deep BALs cutting the emission lines are seen in \ion{C}{iv} and \ion{N}{v}; they extend over 2500 km~s$^{-1}$ and are partially constituted of individual, narrower absorptions. Only the high velocity individual BAL absorptions at $z$ = 2.3229$\pm$0.0005 are seen in \ion{Si}{iv}. Another absorber with narrow lines is only detected in \ion{C}{iv} and \ion{N}{v} at $z$ = 2.3021$\pm$0.0006. Thanks to the spectral resolution of our spectra, we see that the absorption line profiles are nearly identical in the spectra of components A and B. Since it is very unlikely to find the same intrinsic absorbers in two physically distinct quasars, this observation strongly supports the fact that J0818+0601 is actually a gravitationally lensed quasar and not a binary.

A strong intervening narrow absorption system is detected at $z$ = 1.0065$\pm$0.0002 in the spectra of both images A and B, with different intensities. The difference between this intervening system and intrinsic absorptions is clearly seen in the A/B flux ratio. Identified absorption lines of this intervening system are reported in Table~\ref{tab:lines}. The \ion{Mg}{ii} doublet is also detected in absorption at $z$ = 1.0161$\pm$0.0001, but in image~B only. A few additional absorption lines remain unidentified.

Nondetection of a lensing galaxy associated to the A and B components of J0818+0601 led \citet{2016More} to conclude that, if present, the lens should be at a relatively high redshift $z \gtrsim  1$. The metal-line absorption system detected in both images A and B at $z \simeq$ 1.0065 could, therefore, originate from the lensing galaxy. Instead, the system detected at $z \simeq$ 1.0161 might be associated to the lens but this is less likely as it is detected in only one of the images.

\begin{table}
\caption{Intervening absorption lines identified in J0818+0601}
\label{tab:lines}
\centering
\begin{tabular}{ll}
\hline\hline
  Ion & $\lambda_{\rm vacuum}$ \\
\hline
 \ion{Mg}{i}  & 2852.9642 \\   
 \ion{Mg}{ii} & 2803.5310 \\    
 \ion{Mg}{ii} & 2796.3520 \\    
 \ion{Fe}{ii} & 2600.1729 \\   
 \ion{Fe}{ii} & 2586.6500 \\    
 \ion{Fe}{ii} & 2382.7652 \\     
 \ion{Fe}{ii} & 2374.4612 \\     
 \ion{Fe}{ii} & 2344.2139 \\     
  \hline
\end{tabular}
\tablefoot{These lines are detected in both images A and B at redshift $z$ = 1.0065$\pm$0.0002.}
\end{table}

\section{Microlensing in J0818+0601}
\label{sec:mulens}

\begin{figure}
\centering
\resizebox{\hsize}{!}{\includegraphics*{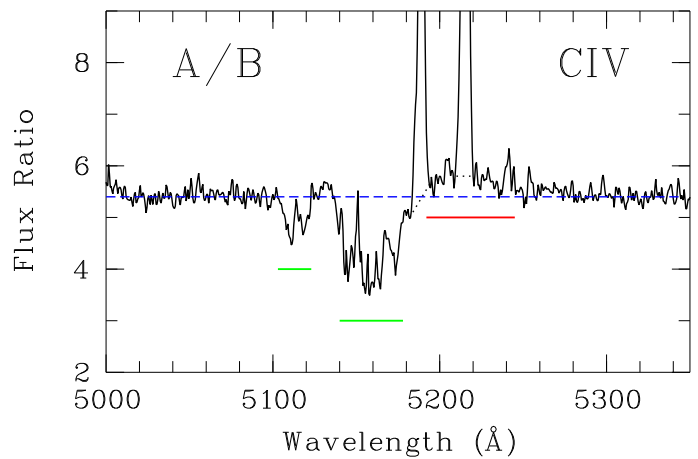}}
\caption{A/B flux density ratio focused on the \ion{C}{iv} spectral region. The blue dashed line represents the adjacent, unabsorbed, continuum ratio. The red line underlines the BEL ratio (interpolated with black dots when contaminated by intervening absorptions), which is significantly higher than the unabsorbed continuum ratio. The green lines underline the flux ratio in the intrinsic absorption lines, which appears significantly lower than the unabsorbed continuum ratio. }
\label{fig:ratiociv}
\end{figure}

If the quasar is only macrolensed, the A/B flux ratio in the line profiles should be identical to the flux ratio in the adjacent continuum. Looking at the region around the \ion{C}{iv} line in Fig.~\ref{fig:ratiociv}\footnote{Since intervening absorption lines are deeper in one image, meaning that the lines of sight to images A and B cross different parts of the intervening absorber, they pop up in the flux ratio.\ However, they do not convey any useful information on the microlensing analysis, they only disturb it.}, we see that this ratio is significantly different between a large part of the BEL and the adjacent continuum, indicating either microlensing or line profile variation with a time delay between images A and B. As discussed in the next section, the latter interpretation can be ruled out. An A/B flux ratio lower in the continuum than in the BEL thus suggests magnification of the continuum source in image~B; indeed, the BELR is less affected by microlensing because its size is much larger than that of the continuum source.

Furthermore, microlensing should similarly affect the absorbed and unabsorbed continua that come from the same source. We then expect that the flux ratio would stay flat across the intrinsic absorption lines, or higher than the adjacent continuum ratio if the absorption lines are partially filled in by the emission lines. This is not observed; the flux ratio is clearly much lower in the absorption lines. A possible interpretation could be selective magnification of a BAL cloud located in front of the continuum source \citep{1993Hutsemekers}. However, in J0818+0601, the BAL cuts the BEL so that the BAL flow must be large enough to cover the BELR. It is therefore unlikely that microlensing resolves inhomogeneities in the BAL flow seen in projection across a continuum source orders of magnitude smaller than the BELR. Instead, this suggests the existence of two sources of continuum that are differently microlensed and absorbed\footnote{Another possibility not related to microlensing would be that the light beams from images A and B intercept different parts of the BAL flow \citep[e.g.,][]{2003Chelouche}. As long as the distance between the quasar accretion disk and the BAL region is smaller than a few tens of parsecs, which is significantly smaller than the distance to the full beam separation, this possibility is unlikely \citep{2015Sluse}. Moreover, considering that the BELR is a background source for the BAL flow, the beam separation occurs at distances 2-3 orders of magnitude larger than when only considering the accretion disk as a source, thus definitely ruling out this interpretation in the case of J0818+0601.}. These continuum sources are expected to have significantly different sizes to be differently microlensed and covered by the BAL flow so as to reproduce the behavior observed in the A/B flux ratio (see Appendix \ref{sec:apb}). Evidence for two sources of continuum is similarly reported in H1413+117 \citep{2015Sluse}.

Finally, as seen in Fig.~\ref{fig:spectra}, the A/B continuum flux ratio shows a slow wavelength dependence over the full wavelength range. This wavelength dependence can be easily explained by differential extinction in the lens galaxy (reddening of image~B). On the other hand, the microlensing effect discussed above can also be a viable explanation depending of the relative contribution of the microlensed and nonmicrolensed continua.

\subsection{MmD spectral decomposition}

\begin{figure}
\centering
\resizebox{\hsize}{!}{\includegraphics*{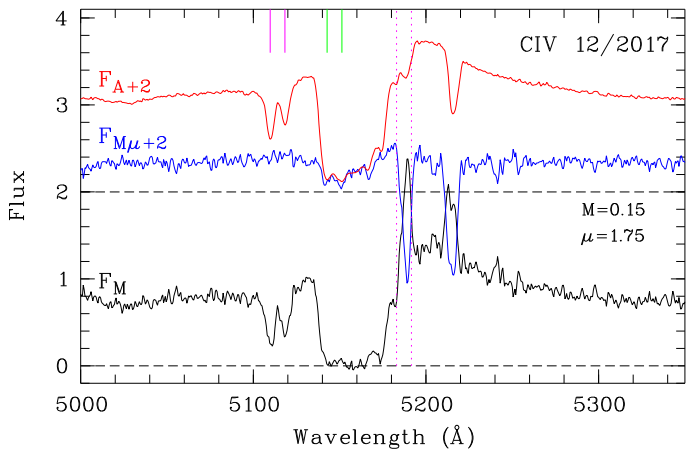}}
\caption{MmD disentangling of the \ion{C}{iv} line profile using the spectra of images A and B of the lensed quasar J0818+0601 secured in December 2017 with the VLT. $F_M$ is the part of the spectrum not affected by microlensing. $F_{M\mu}$ is the part of the spectrum that is micromagnified. $F_A$  = $F_M$ + $F_{M\mu}$ is the spectrum of image~A. These spectra are given in arbitray units.  $F_{M\mu}$ and $F_A$ are shifted upward by two units to enhance visibility. $M$ is the macromagnification factor of image~B relative to image A. $\mu$ is the microlensing magnification factor. The vertical dotted lines indicate the position of the \ion{C}{iv} emission doublet for the systemic redshift $z$ = 2.349. The positions of the intrinsic absorption lines are indicated as in Fig.~\ref{fig:spectra}}
\label{fig:mmdciv}
\end{figure}

\begin{figure}
\centering
\resizebox{\hsize}{!}{\includegraphics*{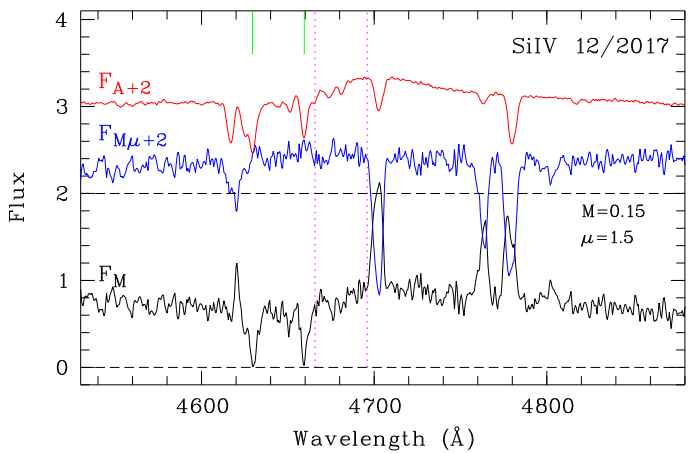}}
\caption{Same as Fig.~\ref{fig:mmdciv}, but for the \ion{Si}{iv} line profile.}
\label{fig:mmdsiiv}
\end{figure}

\begin{figure}
\centering
\resizebox{\hsize}{!}{\includegraphics*{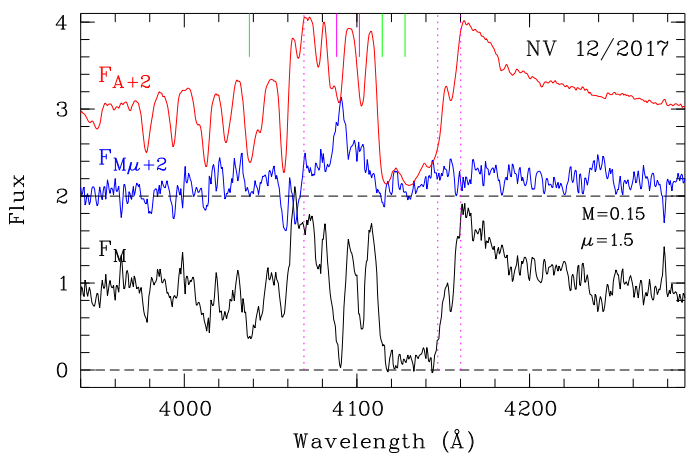}}
\caption{Same as Fig.~\ref{fig:mmdciv}, but for the \ion{N}{v}+Ly$\alpha$ line profile.}
\label{fig:mmdnv}
\end{figure}

\begin{figure}
\centering
\resizebox{\hsize}{!}{\includegraphics*{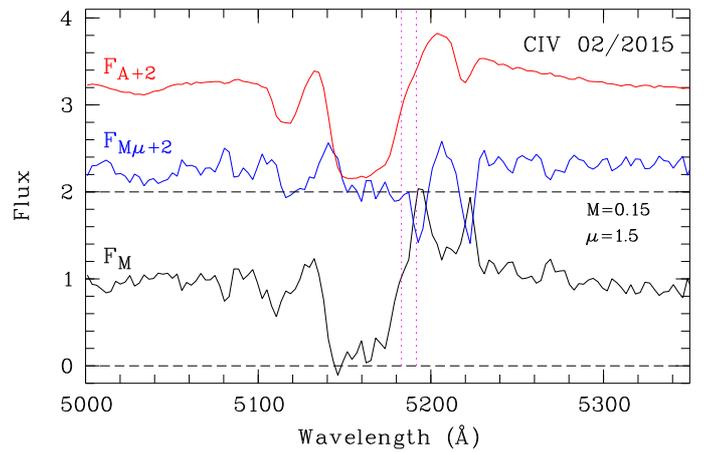}}
\caption{Same as Fig.~\ref{fig:mmdciv}, but for spectra obtained in February 2015 with the SUBARU telescope \citep{2016More}.}
\label{fig:mmdcivsub}
\end{figure}

In order to isolate the part of the spectrum that is microlensed from the nonmicrolensed part, we use in the following the MmD disentangling method developed in \citet{2007Sluse} and \citet{2010Hutsemekers}. We assume that the observed spectra of the lensed images are made of a superposition of a spectrum $F_M$ that is only macrolensed, and of a spectrum $F_{M\mu}$ that is both macro- and microlensed. We then extract the components $F_M$ and $F_{M\mu}$ by using pairs of observed spectra. We first write
\begin{eqnarray} 
F_{\rm B} & =\, & M F_M + M \mu F_{M\mu}  \\ 
F_{\rm A} & =\, & F_M +F_{M\mu}  \; ,
\end{eqnarray} 
where  $F_{\rm A}$ and $F_{\rm B}$  are the observed spectra of images A and B, respectively, $M=M_{\rm B}/M_{\rm A}$ is the macromagnification ratio between images~B and~A, and $\mu$ is the micromagnification factor of image~B. These equations can be rewritten as
\begin{eqnarray} 
F_M \ & = & \frac{-A \;}{A - M} \; \; \left( \frac{F_{\rm B}}{A} - F_{\rm A} \right) \\ 
F_{M\mu} & = & \frac{M}{A - M } \; \; \left( \frac{F_{\rm B}}{M} - F_{\rm A} \right)  \; ,
\end{eqnarray}
where $A = M\mu$.  Up to a scaling factor, $F_M$ only depends on $A$, while $F_{M\mu}$ only depends on $M$. The parameter $A$ is determined as the value for which $F_M(A) \geq 0$ in the (absorbed) continuum, and $M$ is tuned to have most of the emission line appearing in $F_M$ and not in $F_{M\mu}$. We refer to \citet{2015Sluse} for more details in the context of a very similar case.

The results of the MmD spectral decomposition into the part of the spectrum that is only macrolensed, $F_M$, and the part of the spectrum that is also microlensed, $F_{M\mu}$, are illustrated in Figs.~\ref{fig:mmdciv}-\ref{fig:mmdnv} for the spectral regions including the \ion{C}{iv}, \ion{Si}{iv}, and \ion{N}{v}+Ly$\alpha$ lines, respectively, together with the estimated values of $M$ and $\mu = A/M$. Uncertainties on $A$ and $M$ were derived from the range of acceptable decompositions. $F_M(A)$ should be positive everywhere in the absorption lines, but values of $A$ for which $F_M(A)$ is slightly negative remain acceptable given the presence of noise. This defines a small range of $A$ values from which an error was roughly estimated. Similarly, $M$ was tuned to have a flat continuum at the location of the BEL in $F_{M\mu}$ but the presence of noise and the fact that some emission could remain in $F_{M\mu}$ results in a range of acceptable $M$ values. Moreover, to avoid systematics, the decompositions were performed by the three authors independently. We finally estimate $M \simeq$ 0.15 with an error about 5\% for all lines, and $\mu = A/M$ = 1.75$\pm$0.1, 1.5$\pm$0.15, and 1.5$\pm$0.15 from the analysis of the \ion{C}{iv}, \ion{Si}{iv}, and \ion{N}{v} lines, respectively.

Again, we focus on the spectral region around the \ion{C}{iv} line, which gives the most accurate decomposition. The continuum adjacent to the line clearly shows two components, a microlensed one, hereafter $F_{c}$, which is the continuum that appears in $F_{M\mu}$ and a nonmicrolensed one, hereafter $F_{e}$, which is the continuum that is visible in $F_{M}$ together with the BEL. We measured $F_{e}/F_{c} \simeq 2$\footnote{The same decomposition can be obtained assuming a weakly microlensed continuum instead of a nonmicrolensed one at the cost of significantly increasing the flux ratio $F_{e}/F_{c}$.}. Such a separation is the direct consequence that $F_M$ must be positive across the BAL; without the BAL, the existence of two continua would have remained unnoticed \citep{2015Sluse}. The MmD decomposition shows that the BAL flow absorbs these continua with different opacities, completely blocking the nonmicrolensed continuum seen in $F_{M}$. The intrinsic absorber at $z \simeq$ 2.302 only absorbs the nonmicrolensed continuum (in $F_{M}$). We emphasize that the detection of the nonmicrolensed continuum is robust: its suppresssion in $F_{M}$  would require $\mu \simeq 1.2$ instead of $\mu \simeq 1.75$ and result in a strongly negative, unphysical, flux in the BAL. Its significance can also be assessed from the A/B flux ratio, which is definitely much lower in the BAL than in the adjacent continuum (Fig.~\ref{fig:ratiociv}).

With only observations at one epoch, it is not possible to know if the spectral differences between images A and B are due to microlensing or to an intrinsic variation delayed in one image. We have thus applied the MmD to the spectra secured in 2015, by \citet{2016More}, which were kindly provided by the authors. Although small differences can be noticed, a very similar decomposition is obtained, with $M \simeq$ 0.15$\pm$0.01  and $\mu$ = 1.5$\pm$0.2 (Fig.~\ref{fig:mmdcivsub}). Since time delays between the images of double lensed quasars with image separation around 1\arcsec\ are usually no longer than a few months \citep[e.g.,][]{2011Eulaers},  observing the same behavior at an approximately three-year-long interval rules out the possibility that intrinsic variability can explain the spectral differences observed between images A and B, hence strongly supporting the microlensing interpretation.

The MmD spectral decomposition is similar for the \ion{Si}{iv} and \ion{N}{v}+Ly$\alpha$ lines, although it is less clear in the latter case due to lower signal to noise and intervening absorptions. The values of $\mu$ derived for these lines do not differ from the $\mu$ value derived for \ion{C}{iv} within the uncertainties. The macromagnification factor is consistently derived around $M \simeq$ 0.15 for all lines and epochs. This supports the robustness of the results derived from the MmD decomposition.

\subsection{Polarization microlensing}

Spectropolarimetry of images A and B of J0818+0601 shows low polarization without any interesting, significant features in the polarized spectrum. We then integrated the spectra over the 3900-6000 \AA\ spectral range to increase the signal to noise ratio before recomputing the polarization quantities. We finally derived the ``white light'' polarizations $p_A$ = 0.77$\pm$0.03\% with $\theta_A$ = 97$\pm$1\degr\ and  $p_B$ = 0.31$\pm$0.10\% with $\theta_B$ = 178$\pm$9\degr\ for images A and B, respectively. The polarization of image~A is typical of high-ionization BAL quasars \citep{1998Hutsemekers}.

The difference in the polarization observed between the two images can be explained by microlensing in image~B. Indeed, a change of the relative contribution of two continua differently polarized, due to microlensing magnification of one of them, can explain different polarization degrees and the rotation of the polarization position angle \citep{2015Hutsemekers}. On the other hand, a polarization difference between images A and B can also occur due to differential interstellar polarization in the lens galaxy, especially since differential extinction cannot be excluded from the observed A/B continuum flux ratio (Fig.~\ref{fig:spectra}). Monitoring the polarization of both images can help identify the right scenario.

\section{Implications for the quasar structure}
\label{sec:discu}

\begin{figure}
\centering
\resizebox{\hsize}{!}{\includegraphics*{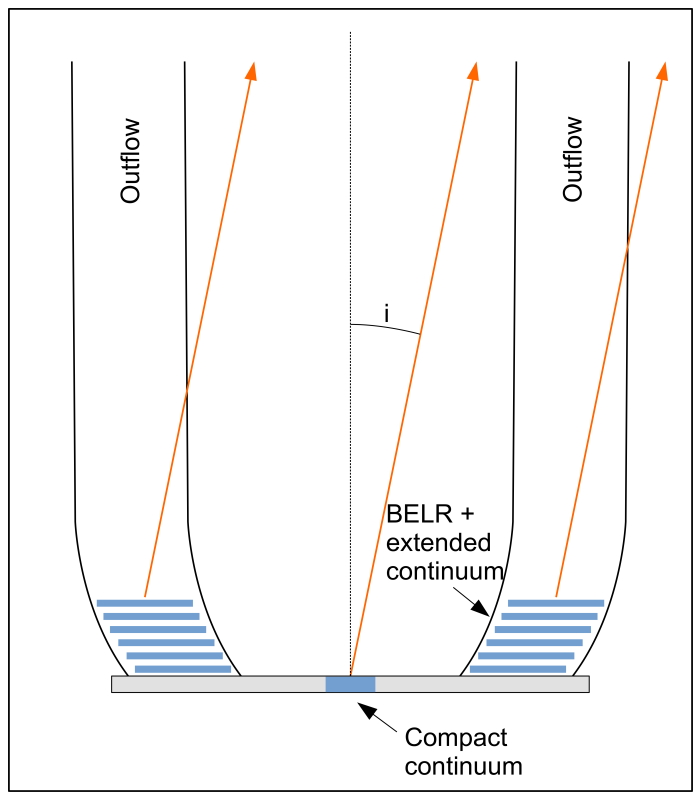}}
\caption{Cartoon (not to scale) representing a possible location of the extended continuum source in the central region of J0818+0601. The compact UV continuum source is located close to the center of the accretion disk. An outflow originates from the outer accretion disk. The BELR and the extended continuum source (hatched region) are located at the base of the outflow. Lines of sight to the observer, with an inclination $i$, are indicated. The compact continuum source is assumed to be microlensed, which is not the case for the BELR and extended continuum.}
\label{fig:diskwind}
\end{figure}

Thanks to microlensing in one of the lensed images and to the presence of a BAL flow, we unveiled two sources of continuum in the quasar J0818+0601 from both the analysis of the A/B flux ratio and the MmD method. A similar finding was previously made in the lensed BAL quasar H1413+117, thus strengthening the case. The microlensed continuum is naturally expected to come from a compact source, likely the accretion disk, and the nonmicrolensed one from a more extended region. Because this extended continuum is detected in the rest-frame UV and absorbed by the BAL flow, it does not come from the host galaxy but from the quasar inner regions. Its size should at least be larger than the Einstein radius of the system, $\eta_0$ = 7.6 lt-days computed for an average microlensing star mass of 0.3 M$_{\odot}$ and assuming the lens galaxy is at $z$ = 1.0065.

Several scenarios explaining the presence of an extended source of continuum are discussed in \citet{2015Sluse}. In the case of H1413+117, the nonmicrolensed extended continuum is preferentially interpreted as light scattered in the neighborhood of the central engine, possibly in a polar flow \citep{2015Hutsemekers}. Even so, the model of a false continuum coming from reflection onto a population of cold, thick, clouds \citep{2012Lawrence} has to be considered. Evidence for the existence of a diffuse or extended continuum was also discussed in the framework of reverberation mapping experiments carried out for low redshift active galactic nuclei. In particular, models involving a diffuse continuum emitted from the BELR itself \citep{2001Korista,2018Lawther,2019Goad}, from its launching region \citep{2019Chelouche}, or from the base of a disk-wind \citep{2019Dehghanian}, could also potentially explain the extended source of continuum found in J0818+0601.  On top of that, to reach the flux ratio $F_{e}/F_{c} \simeq 2$, it might be necessary to add some extinction along the line of sight to the compact continuum source. An accurate measurement of the wavelength dependence of the continuum flux ratio $F_{e}/F_{c}$ would help to constrain these scenarios.

The black absorption in the \ion{C}{iv} BAL ($F_{M}$ in Fig.~\ref{fig:mmdciv}) indicates that the BAL flow fully covers the extended continuum source and the BELR in J0818+0601. In the framework of disk-wind models, the BELR is located in the wind, above the outer accretion disk.  Assuming that the extended continuum source is cospatial with the BELR, the dense wind launched from the disk could cover both regions as illustrated in Fig.~\ref{fig:diskwind}. The line of sight to the compact continuum source should be sufficiently separated to allow for differential microlensing. Moreover it should cross less opaque streamlines in the outflow (cf. $F_{M\mu}$ in Fig.~\ref{fig:mmdciv}), which is more easily explained if the system is seen under low inclination. The fact that the \ion{C}{iv} and \ion{N}{v} BALs have a nonzero ($\sim$900 km s$^{-1}$) onset velocity\footnote{The onset velocity is measured in the BAL troughs as the lowest velocity at which the flux first reaches 50\% of the adjacent continuum flux. The velocity is measured with respect to the red line of the emission line doublets, assuming the systemic redshift $z$ = 2.349.} may be the signature of a disk-wind \citep{1995Murray}.  The \ion{Si}{iv} line, on the other hand, only shows narrower individual components with a velocity of $\sim$2200 km s$^{-1}$ , suggesting ionization stratification.  Interestingly, along the line of sight to the compact continuum source ($F_{M\mu}$ in Figs.~\ref{fig:mmdciv} and~\ref{fig:mmdsiiv}), the \ion{C}{iv} BAL seems to have a higher onset velocity ($\sim$1300 km s$^{-1}$), while \ion{Si}{iv} absorption is absent. We finally note that a polar wind component might be needed to account for polarization properties specific to BAL quasars \citep{2004Lamy,2010Hutsemekers,2015Hutsemekers}.

The higher velocity absorber at $z \simeq$ 2.302 ($\sim$4200 km s$^{-1}$), which is only seen along lines of sight to the extended continuum ($F_{M}$ in Figs.~\ref{fig:mmdciv} and~\ref{fig:mmdnv}), shows evidence for partial covering. We estimate a covering factor of $\sim$70\% from the \ion{C}{iv} doublet ratio using formulae given in \citet{2004Hutsemekers}. Finally, although the data are of lesser quality, the spectra obtained in 2015 (Fig.~\ref{fig:mmdcivsub}) might indicate some time variability (cf. a possible absorption in $F_{M\mu}$ in 2015).

\section{Conclusions}
\label{sec:conclu}

First, our new spectroscopic observations indicate that J0818+0601 is actually gravitationally lensed, and not a binary quasar. Detection of an absorption system at $z$ = 1.0065$\pm$0.0002 might reveal the lensing galaxy.

The differential microlensing magnification observed in one image brings out two, spatially separated, sources of continuum. The extended, nonmicrolensed continuum contributes to two thirds of the total continuum emission. J0818+0601 is the second BAL quasar in which an extended source of continuum is found. We find possible evidence for polarization microlensing, suggesting that the two continua might be differently polarized. Disentangling the microlensed and nonmicrolensed parts of the spectra also provides constraints on the BAL flow. In particular, we find that the wind has a nonzero onset velocity along the line of sight to the observer and is  ionization stratified.

Monitoring the spectrum and polarization of images A and B of J0818+0601 would help constrain the various models for the extended continuum and the outflow. In particular, the wavelength dependence of the ratio $F_{e}/F_{c}$ over a wide spectral range might reveal whether scattering is responsible for the extended continuum. The detection of polarization variations would support polarization microlensing and the possible existence of two differently polarized continua. The variability of the absorption features in either the compact or the extended continuum would provide information on the structure of the outflow.

\begin{acknowledgements}
  
We thank A.~More and M.~Oguri for kindly providing us with the spectra published in \citet{2016More}. We thank the referee for comments that help to improve the manuscript. P.K. thanks the University of Li\`ege, Belgium, for an ERASMUS fellowship.

\end{acknowledgements}

\bibliographystyle{aa}
\bibliography{references}

\begin{appendix}
\section{Data reduction and spectral extraction}
\label{sec:apa}

The data have been reduced using the ESO-REFLEX pipeline \citep[v.2.9.1.;][]{2013Freudling}, which implements standard spectroscopic reduction steps (i.e., bias subtraction, flat fielding, geometric correction, and wavelength calibration) for each 2D spectrum constituting the spectropolarimetric data set. A polynomial of order 4 was used for the spectral wavelength calibration, and an order 2 was employed for the spatial curvature correction. The final set of spectra is resampled on a 0.75\AA/pixel grid.  We paid particular attention to potential sources of systematic errors in the reduction, which could induce a spurious polarization signal. In particular, we verified the flat field correction at the position of the spectra on the CCD in the ordinary and extraordinary beams. We find the two corrections to be identical to 1.000$\pm$0.004 (68\% confidence level).  On the other hand, the wavelength calibration is found to be accurate to better than 0.2 pixel (root mean square), without systematic differences between the two beams, except for $\lambda < 3900$\AA. This wavelength range is, however, not considered in our analysis. Finally, the spectra of the individual lensed images were extracted by fitting a sum of 2 Moffat profiles along the spatial direction, fixing their separation to 8.85$\pm$0.02 pixels = 1.115$\pm$0.003\arcsec, that is, the median value of the separation found when leaving the separation as a free parameter. During this process, we also fixed the poorly constrained $\beta$ parameter of the Moffat profile to $\beta=4.5$ (i.e., a slightly lower value than expected from atmospheric theory; \citealt{2001Trujillo}). After extraction, the residuals at the location of image~B are lower than 2\% of the flux of B over the whole wavelength range. Since we are only interested in relative fluxes, no absolute flux calibration was performed. In total, 16 spectra (two orthogonal polarizations, four half-wave plate angles, and two epochs) were obtained for each quasar image. Data obtained on December 24 and December 25 were averaged in order to obtain eight spectra per image, which were subsequently used to compute the polarization quantities.

\section{Differential microlensing in the A/B flux ratio}
\label{sec:apb}

In order to interpret the A/B flux ratio observed in the spectral region around the \ion{C}{iv} line (Fig.~\ref{fig:ratiociv}), we consider an  emission and absorption line profile where both the continuum and the emission line are absorbed. If $M$ and $\mu$ are respectively the macro- and micromagnification factors, we write for images B and A:
\begin{eqnarray} 
F_{\rm B}(\lambda) & = & M \, F_{\rm em} \, e^{-\tau_{\rm em}} +  M \, F_{e} \, e^{-\tau_{e}} + M \mu \, F_{c} \, e^{-\tau_{c}} \\ 
F_{\rm A}(\lambda) & = & F_{\rm em} \, e^{-\tau_{\rm em}} + F_{e} \, e^{-\tau_{e}} + F_{c} \, e^{-\tau_{c}}
,\end{eqnarray} 
where $F_{e}(\lambda)$ represents the flux from an extended source of continuum that is only macrolensed as the emission line flux $F_{\rm em}(\lambda)$, and $F_{c}(\lambda)$ is the flux from a compact source of continuum that is also microlensed. These components are absorbed with different opacities $\tau_{e}(\lambda)$, $\tau_{\rm em}(\lambda)$, and $\tau_{c}(\lambda)$, respectively. If $\mu > 1$, we have $1/(M\mu) \leq F_{\rm A}/F_{\rm B} \leq 1/M$. If $F_{e}$ = 0, the minimum value of $F_{\rm A}/F_{\rm B}$ is reached in the continuum, out of the line profile, that is, where $ F_{\rm em} = 0$ and $\tau_{c} = 0$. In this case, $F_{\rm A}/F_{\rm B}$ in the profile is larger or equal to $F_{\rm A}/F_{\rm B}$ in the adjacent continuum. If $F_{e} > 0$, then $F_{\rm A}/F_{\rm B} > 1/(M\mu)$ in the continuum, out of the line profile. The lower limit on  $F_{\rm A}/F_{\rm B}$ can be reached in the absorption line if $\tau_{\rm em} = \tau_{e} = \infty$.  When  $F_{\rm A}/F_{\rm B}$ in the absorption profile is lower than $F_{\rm A}/F_{\rm B}$ in the adjacent continuum, this is thus an indication that  $F_{e} > 0$, that is, it signals the presence of a nonmicrolensed continuum.

\end{appendix}

\end{document}